\documentclass[twocolumn,showpacs,preprintnumbers,amsmath,amssymb,epsfig,widetext]{revtex4}

\usepackage{graphicx}
\usepackage{dcolumn}
\usepackage{bm}
\usepackage{epsfig}

\def\fun#1#2{\lower3.6pt\vbox{\baselineskip0pt\lineskip.9pt
  \ialign{$\mathsurround=0pt#1\hfil##\hfil$\crcr#2\crcr\sim\crcr}}}
\def\simgt{\mathrel{\lower0.6ex\hbox{$\buildrel {\textstyle >}
 \over {\scriptstyle \sim}$}}}
\def\simlt{\mathrel{\lower0.6ex\hbox{$\buildrel {\textstyle <}
 \over {\scriptstyle \sim}$}}}

\input epsf

\newcommand{\hompc}{\,h\,{\rm Mpc}^{-1}}
\newcommand{\mpcoh}{\,h^{-1}\,{\rm Mpc}}

\def\be{\begin{equation}}
\def\ee{\end{equation}}
\def\ba{\begin{eqnarray}}
\def\ea{\end{eqnarray}}

\def\nn{\nonumber}

\begin{document}

\preprint{}

\title{Reconstructing the History of Structure Formation using Redshift Distortions}

\author{Yong-Seon Song and Will J. Percival}
\affiliation{Institute of Cosmology $\&$ Gravitation, Dennis Sciama building, 
University of Portsmouth, Portsmouth, PO1 3FX, UK }

\date{\today}

\begin{abstract}
  Measuring the statistics of galaxy peculiar velocities using
  redshift-space distortions is an excellent way of probing the
  history of structure formation. Because galaxies are expected to act
  as test particles within the flow of matter, this method avoids
  uncertainties due to an unknown galaxy density bias. We show that
  the parameter combination measured by redshift-space distortions,
  $f\sigma_8^{\rm mass}$ provides a good test of dark energy models,
  even without the knowledge of bias or $\sigma_8^{\rm mass}$ required
  to extract $f$ from this measurement (here $f$ is the logarithmic
  derivative of the linear growth rate, and $\sigma_8^{\rm mass}$ is
  the root-mean-square mass fluctuation in spheres with radius
  $8h^{-1}$Mpc). We argue that redshift-space distortion measurements
  will help to determine the physics behind the cosmic acceleration,
  testing whether it is related to dark energy or modified gravity,
  and will provide an opportunity to test possible dark energy
  clumping or coupling between dark energy and dark matter. If we can
  measure galaxy bias in addition, simultaneous measurement of both
  the overdensity and velocity fields can be used to test the validity
  of equivalence principle, through the continuity equation.
\end{abstract}



\maketitle

\section{introduction}

Observations of the accelerating nature of the Universe show that
there is fundamental physics at work that we do not understand
\cite{riess98,perlmutter99}. Many possibilities have been postulated
including a new form of the vacuum which is not present in the
contemporary high energy physics, or a modification of gravity which
would revolutionize our understanding of space and time. Discovering
which mechanism is correct is one of the key challenges for 21st
century science.

The detection of acceleration was obtained by using supernovae as
standard candles, and therefore relies on measuring the cosmological
geometry. The physical process causing the acceleration could also
affect structure formation, which provides a complementary way of
distinguishing between models. In particular, models in which general
relativity is unmodified have different Large-Scale Structure
(hereafter LSS) formation timescales compared with Modified Gravity
(hereafter MG) models~\cite{dvali00,carroll03}. Because of this, many
previous studies have considered testing the signature on LSS
formation predicted by MG models~\cite{Lue:2003ky,ishak05,jain07}.
Direct observations of LSS growth as traced by galaxies are of limited
value because galaxies are not expected to be simple tracers of the
underlying matter density field, although we return to this point
later.

Maps of galaxies where distances are measured from spectroscopic
redshifts show anisotropic deviations from the true galaxy
distribution. These differences arise because galaxy recession
velocities include components from both the Hubble flow and peculiar
velocities from the motions of galaxies in comoving space.  Although
these ``redshift-space distortions'' are a nuisance when trying to
reconstruct the true distribution of galaxies, they provide a
mechanism to measure the build-up of structure, which drives these
peculiar velocities on large-scales. 

In linear theory, and in the absence of bias, a distant observer
should expect a multiplicative enhancement of the overdensity field
$\delta$ that is proportional to $1+f\mu^2$, where $f$ is the
logarithmic derivative of the linear growth rate, and $\mu$ is the
cosine of the angle to the line-of-sight \cite{kaiser87}. With a local
linear bias, the real-space galaxy density field is affected, while
the peculiar velocity term is not, so the multiplicative factor is
changed to $1+\beta\mu^2$, where $\beta\equiv f/b$. Because of the
$\mu$ dependence, this information can be extracted from galaxy
redshift surveys, and a number of methods and their application have
previously been considered. Analyses have been undertaken using the
2degree-Field Galaxy Redshift Survey (2dFGRS) \cite{colless03},
measuring redshift-space distortions in both the correlation function
\cite{peacock01,hawkins03} and power spectrum after decomposing into
an orthonormal basis of spherical harmonics and spherical Bessel
functions \cite{percival04}. Using the Sloan Digital Sky Survey
\cite{York00}, an Eigenmode decomposition has been performed to
separate real and redshift-space effects
\cite{tegmark04,tegmark06}. In a recent paper, these low redshift
analyses were extended to $z\simeq1$ \cite{guzzo08} using the
VIMOS-VLT Deep Survey (VVDS) \cite{lefevre05,garilli08}.  In addition
to measuring $\beta$ at $z=0.8$, this work has pushed explicitly the
idea of using large-scale peculiar velocities for constraining models
of cosmic acceleration.  These data, and the resulting cosmological
constraints, are considered further in Section~\ref{sec:current}. In
particular we argue that it would be better to present results in
terms of $b(z)\sigma_8(z)$ and $f(z)\sigma_8(z)$, rather than $\beta$
for local linear bias models. We show in
Section~\ref{sec:pec_vel_meas} that bias-independent constraints on
$f(z)\sigma_8(z)$ are able to discriminate between some models of
acceleration as well as $f(z)$, which is commonly extracted from
$\beta$ by applying an independent (and difficult) measurement of
bias. Galaxy bias measurements tend to have the same fractional error
as the redshift-spce distortion measurements: for example comparing
redshift-space distortion results from the 2dFGRS\cite{hawkins03} with
bias measurements from measurements of the 3-pt
function\cite{verde02}.

In addition to a direct measurement of $f(z)\sigma_8(z)$,
redshift-space distortion measurements can be used to test diverse
aspects of LSS, as proposed by Song $\&$ Koyama~\cite{song08a}:
geometrical perturbations can be reconstructed from the evolution
history of peculiar velocities.  With the assumption of an additional
measurement of galaxy bias, the continuity equation can be tested, and
anisotropic stress can be constrained.  Those diverse tests strengthen
our power to constrain theoretical models, and are considered further
in Section~\ref{sec:further_tests}. Before we do this, we first review
the physics that we hope to test using peculiar velocities
(Section~\ref{sec:models}), and then consider the measurements
themselves and what they can directly tell us about structure
formation (Section~\ref{sec:pec_vel_meas}).

\section{Linear Structure Formation} \label{sec:models}

\subsection{Basic equations}

In this section, we briefly review the standard derivation of the
dynamics of the density fluctuations and their associated peculiar
velocities in a Friedman universe. We will contrast this against
non-standard models in later sections. In the Newtonian gauge, the
perturbed metric describing local gravitational instability of the
energy-momentum density fluid is given by
\begin{equation}
 ds^2=-(1+2\Psi)dt^2+a^2(1+2\Phi)dx^2\,,
\end{equation}
where scalar perturbations are dominant over vector or tensor
perturbations. The Newtonian force $\Psi$ sources the dynamics of the
perturbed fluids, while the curvature perturbation $\Phi$ measures the
local energy density fluctuations. In the epoch of non-relativistic
particle domination, in standard GR models, $\Phi$ is identical to
$-\Psi$, and there is no anisotropic stress.

The spatial variation of density fluctuations is expressed by the
density contrast $\delta_X =\delta\rho_X(t,{\bf x})/\bar\rho_X(t)$
where $X$ denotes the specific fluid that we are considering. In the
approximation of negligible irrotational flow, the divergence of the
peculiar velocity ${\bf v}_X$, $\theta_X\equiv{\bf \nabla}\cdot {\bf
  v}_X$ can be used to describe the fluid motion. The dynamics of
scalar perturbations of the fluid $X$ is well described by energy
momentum fluctuations $\delta_X$ and $\theta_X$, and the corresponding
metric perturbations $\Psi$ and $\Phi$.

The conservation equation $\nabla_{\mu} T^{\mu}_{\nu}=0$ gives the set
of equations that describe the dynamics of fluid $X$,
\begin{eqnarray}
\frac{d\delta_X}{dt}&=&-(1+w_X)\frac{\theta_X}{a}-3H\frac{\delta p_X}{\rho_X}
+3Hw_X\delta_X\,,\label{eq:conteq_X}\\
\frac{d\theta_X}{dt}&=&-H(1-3w_X)\theta_X-\frac{dw_X/dt}{1+w_X}\theta_X\nn\\
&&+\frac{k^2}{a}\left(\frac{\delta p_X}{\rho_X}\frac{1}{1+w_X}
-\sigma_X+\Psi\right)\,,\label{eq:Euler_X}
\end{eqnarray}
where $w_X$ is the equation of state, $\delta p_X$ is the perturbed
pressure and $\sigma_X$ is the anisotropic stress.  The continuity
equation, Eq.~(\ref{eq:conteq_X}), states conservation of local density.
The Euler equation, Eq.~(\ref{eq:Euler_X}), represents the conservation
of local energy momentum, and describes dynamics of perturbed fluids
sourced by $\Psi$.

The curvature perturbation $\Phi$ is constrained to the local inhomogeneity
via the Poisson equation,
\begin{equation}\label{eq:Poisson_X}
 k^2\Phi=4\pi G_N a^2 \rho_X\left(\delta_X+3aH\frac{\theta_X}{k^2}\right)\,.
\end{equation}
These equations completely determine the dynamical evolution of LSS,
within a given expansion history $H$.

For models based on general relativity with a standard dark energy
that does not clump on small scales, the scale and time dependence of
the evolution of perturbations are separable in the matter dominated
regime. Both $\delta_X$ and $\theta_X$ are uniquely determined by the
expansion history $H$. Thus we are able to trace the evolution of
structure formation using observations of either $\delta_X$ or
$\theta_X$ in these models.

\subsection{Dark energy model without clumping}\label{sec:DE}

In standard dark energy models (hereafter sDE), the cosmic expansion
is accelerated by introducing a homogeneous dark energy component into
the Friedman equation, which then predicts an expansion rate $H$,
\begin{equation}
H^2=H_0^2\left[\frac{\Omega_b}{a^3}+\frac{\Omega_c}{a^3}
+\frac{\Omega_{DE}}{a^{3(1+w_{DE})}}\right]\,,
\end{equation}
where `$b$' denotes baryon, `$c$' denotes cold dark matter.

The incoherence between baryons and CDM perturbations caused by
acoustic waves in the early universe is removed at redshifts
$z\simgt10$ \cite{eisenstein06}, and can therefore be ignored in our
analysis. We also assume a negligible cosmological neutrino density,
consistent with observations (e.g. \cite{elgaroy02}). Following
these approximations, all matter inhomogeneities can be assumed to
coherently evolve with those in the CDM. Thus we can treat all matter
as a single fluid denoted by `$m$'. Eq.~(\ref{eq:conteq_X}) and
Eq.~(\ref{eq:Euler_X}) for matter fluid `$m$' are,
\begin{eqnarray}
\frac{d\delta_m}{dt}+\frac{\theta_m}{a}&=&0
 \label{eq:conteq_mDE}\\
\frac{d\theta_m}{dt}+H\theta_m&=&\frac{k^2}{a}\Psi\,.
 \label{eq:Euler_mDE}
\end{eqnarray}
Since we are assuming that the dark energy is homogeneous on the
scales of interest, the metric perturbations can be simply related to
the total matter-energy fluctuations,
\begin{equation}\label{eq:Poi_DE}
 k^2\Phi=4\pi G_N a^2 \rho_m\delta_m\,.
\end{equation}
We use $\Lambda$CDM model as an example of sDE with the cosmological 
parameters $w_c(\Omega_ch^2)=0.11$, $w_b(\Omega_bh^2)=0.021$ and $h=0.72$.

The combination of Eq.~(\ref{eq:Poi_DE}), the anisotropic stress relation
$\Phi=-\Psi$, and the coupled equations of Eq.~(\ref{eq:conteq_mDE}) and
Eq.~(\ref{eq:Euler_mDE}) provide a unique solution for the formation of
cosmological structure. See \cite{percival05} for a review of methods
to solve these equations.

\subsection{Modified Gravity: DGP}\label{sec:DGP}

The cosmic acceleration may arise from a modification of gravity on
cosmological scales as in the DGP model \cite{dvali00}.  In this model,
we live on the $(3+1)-$dimensional brane which is embedded in an
infinite Minkowski bulk.  The weakened gravity at cosmological scales
induces the cosmic acceleration without introducing dark energy.  The
expansion history of DGP model is determined by the usual
matter-energy density and the crossover scale defined as the ratio of
5-dimensional to 4-dimensional Planck mass scales
$r_c=M_{pl}^{(4)2}/2M_{pl}^{(5)3}$.
\begin{equation}
H^2-\frac{H}{r_c}=\frac{8\pi G_N}{3}\rho_m\,.
\end{equation}
As an example of DGP models, we take the same $w_c$ and $w_b$ as in
our $\Lambda$CDM model discussed in section~\ref{sec:DE}, but use a
different $h=0.80$ in order to provide a nearly identical $H(a)$.

In general the equations of motion of linear perturbations in DGP is
not closed without solving the dynamic equation of propagation through
the bulk. But in the quasi static limit $k/aH\gg 1$, where the
contribution of bulk gradient is negligible, the solution can be
derived from dynamic equations on the brane. In this regime, the
perturbed potentials in DGP become modified,
\begin{eqnarray}
 k^2\Phi&=& 4\pi G_N \left(1-\frac{1}{3\beta_{DGP}}\right) a^2 \rho_m\delta_m\,,\\
 k^2\Psi&=&-4\pi G_N \left(1+\frac{1}{3\beta_{DGP}}\right) a^2 \rho_m\delta_m\,,
\end{eqnarray}
where 
\begin{equation}
\beta_{DGP}=1-2r_cH\left(1+\frac{\dot H}{3H^2}\right)\,.
\end{equation}
The effective Newtonian constant in Poisson equation $k^2\Phi=4\pi
G_{eff}(a) a^2 \rho_m\delta_m$ becomes
$G_{eff}(a)=G_N(1-1/3\beta_{DGP})$, and a non-trivial anisotropic
stress is introduced
\begin{equation}\label{eq:pi_DGP}
\frac{\Phi}{\Psi}=\frac{1-3\beta_{DGP}}{1+3\beta_{DGP}}\,,
\end{equation}
while there are no changes in the continuity equation and 
the Euler equation~\cite{Lue:2003ky,koyama05,sawicki05}.

The differences between DGP and sDE models lead to distinct evolution
of $\delta_m$ and $\theta_m$ even for models with identical expansion
histories. Coupling measurements of either $\delta_m$ or $\theta_m$
with observations of the geometrical evolution has the potential to
distinguish DGP from sDE models. If we are able to measure both
$\delta_m$ and $\theta_m$ simultaneously, then the non-trivial
anisotropic stress can be measured, which also distinguishes DGP and
sDE models.

\subsection{Dark energy model with clumping}

If the dark energy can support long-lived fluctuations, then the
galaxies will trace the total density fluctuations $\delta_T$ rather
than those just in the matter $\delta_m$. In these clumping dark
energy models (hereafter cDE), the baryons will fall into the
potential wells created by both the dark matter and energy. Given the
different physical behaviour of dark matter and dark energy $\delta_T$
will not be a simple linear function of $\delta_m$. The Poisson
equation with dark energy clumping will be given by
\begin{equation}
 k^2\Phi=4\pi G_N a^2 \rho_T\delta_T\,.
\end{equation}
where $\rho_T\delta_T=\rho_m\delta_m+\rho_{DE}\delta_{DE}$~\cite{Kunz:2006ca}.

When we have multiple components with different equations of state,
the peculiar velocities of the different components will not, in
general, be the same at any spatial location. Consequently, we will
measure $\theta_m$ from galaxy redshift-space distortions, rather than
$\theta_T$. The continuity equation holds for matter and DE
separately, although it does not hold if we mix components. So, for
example,
\begin{eqnarray}
\frac{d\delta_T}{dt}+\frac{\theta_m}{a}\neq 0\,,
 \label{eq:DEGRcont1}
\end{eqnarray}
in general. If we can measure $\delta_T$ and $\theta_m$, we can
distinguish between DGP and cDE by using
Eq.~(\ref{eq:DEGRcont1}).  For example, if the contribution from dark
energy fluctuations are accidentally identical to the effect of
$G_{eff}$ of DGP,
\begin{equation}
G_{eff}(a)=G_N \frac{\rho_T\delta_T}{\rho_m\delta_m}\,,
\end{equation}
and the anisotropic stress causes a similar change to $\Psi$ in
Eq.~(\ref{eq:pi_DGP}) of DGP, then there is no difference in structure
formation as measured by the evolution of $\theta_m$. However,
if we observe $\delta_T$, then we can break the consistency of
Eq.~(\ref{eq:DEGRcont1}), leading to a possible method for
distinguishing between these models. 

\subsection{Interacting dark energy model}

Current observations allow a coupling between dark matter and dark
energy~\cite{Amendola:1999er,Amendola:2003wa}, although the coupling
between baryon and dark energy is strongly limited by current
experimental constraints~\cite{Hagiwara:2002fs}.  For a model with
interacting dark matter and dark energy (hereafter IDE), the set of
background continuity equations are,
\begin{eqnarray}
  \frac{d\rho_b}{dt} + 3H\rho_b&=&0\nn \\
  \frac{d\rho_c}{dt} + 3H\rho_c&=&
  -\gamma\rho_c\frac{d\phi}{dt}\nn \\
  \frac{d\rho_v}{dt} + 3H(\rho_v+p_v)&=& \gamma\rho_c\frac{d\phi}{dt}\,,
\end{eqnarray}
where $\gamma$ is an arbitrary coupling constant. A positive coupling constant
represents decay from dark matter to dark energy, and a negative
coupling constant represents decay from dark energy to dark matter.
The equation of motion of the scalar field is given by
\begin{equation}
  \frac{d^2\phi}{dt^2} +3H\frac{d\phi}{dt}+\frac{dV}{d\phi}=\gamma\rho_c\,.  
\end{equation}

The conservation equation leads to a set of dynamic equations for
baryons and for dark matter. For baryons, there is no signature from
the coupling in either the continuity equation or the Euler
equation. For the dark matter, the coupling influences the dynamics,
\begin{eqnarray}
  \frac{d\delta_c}{dt}+\frac{\theta_c}{a}&=&0\\
  \frac{d\theta_c}{dt}+H\theta_c-\gamma\frac{d\phi}{dt}\theta_c
   &=&\frac{k^2}{a}\Psi-a\gamma^2\rho_c\delta_c\,.  
\end{eqnarray}
The continuity equation still holds because the creation/destruction
rate of dark matter is proportional to its current density, but the
coupling between dark matter and dark energy changes the Euler
equation.

The interaction between dark matter and dark energy modifies the
dynamics of dark matter, so two test particles of baryon and dark
matter placed in the same force field will respond differently. There
is no change in the matter continuity equation but, in general, we
will measure the velocities of baryonic material, while we observe
galaxies in the dark matter potential wells, so measure matter
overdensities. The observed continuity equation will then be broken in
the presence of coupling between dark matter and dark
energy~\cite{song08b}.

\section{Measuring peculiar velocities} \label{sec:pec_vel_meas}

If we map cosmological structure by translating observed galaxy
redshifts to distances assuming that they are cosmological in origin
then peculiar galaxy velocities are misinterpreted, leaving an
anisotropic galaxy distribution. For pairs of galaxies with large
separation, the peculiar velocities can tell us about the formation of
large-scale structure\cite{kaiser87}. On small scales, decoherent
peculiar velocities cause Fingers-of-God (FOG), stretching compact
structures along the line-of-sight. These distortions depend on the
structure of halos and any cosmological information is difficult to
distinguish from halo properties. The redshift-space power spectrum
$P_g^s({\bf k})$ of a galaxy redshift survey is commonly modelled
\cite{kaiser87,pd94} as
\begin{equation}
 P_{gg}^s({\bf k}) = \left[ P_{gg}({\bf k})
                    + 2\mu^2P_{g\theta_g}({\bf k})
                    + \mu^4P_{\theta_g\theta_g}({\bf k})\right]
                    F\left(k^2\sigma_v^2(z)\mu^2\right),
 \label{eq:pk_s1}
\end{equation}
where $\mu=k_\|/k$ is the cosine of the angle of the ${\bf k}$ vector
to the line of sight. $P_g$, $P_{\theta_g\theta_g}$, and
$P_{g\theta_g}$ are the real space auto-power spectra of galaxies and
$\theta_g$, and the cross power spectrum of galaxy-$\theta_g$
fluctuations, respectively. $\sigma_v$ and $F$ determine the
non-linear velocity distribution of galaxies in collapsed
structures. It is common to assume an exponential model for the
pairwise peculiar velocities, so
$F(k^2\sigma_v^2(z)\mu^2)=(1+k^2\sigma_v^2(z)\mu^2)^{-1}$, although
Gaussian models have also been considered \cite{pd94}. Even in the
distant-observer limit, the usefulness of this equation is limited
because Eq.~(\ref{eq:pk_s1}) is not physically motivated
\cite{scoccimarro04}. Additionally, $\sigma_v$ is expected to be a
function of halo mass and redshift. Consequently, in order to exploit
the precision available from future surveys, it may become necessary
to model the exact behaviour using simulations
\cite{hatton98,tinker05,pw08}. In this paper, we assume that
Eq.~(\ref{eq:pk_s1}) holds, and that any significant deviations can be
accurately modelled prior to the new data sets becoming available.

A concern is that the distribution of galaxy pair-velocities on large
scales might not match the distribution of velocities in the matter
field. For example, the velocity power spectrum for peaks in a density
field does not match that of the mass \cite{regos95,percival08}, even
if they are the same locally. In this paper, we follow the standard
ansatz that this bias is small (e.g. \cite{jain07}) and assume that
$\theta_g\simeq \theta$. Current simulations indicate that this might
cause a 10\% systematic error on scales 10--200$\hompc$
\cite{hatton98,tinker05}, although we are confident that this can be
accurately simulated, so will only affect future measurements at a
lower level. In the following we therefore drop the subscripts $g$ and
$m$ from $\theta$.

We ignore any cosmological information in $F(k^2\sigma_v^2(z)\mu^2)$,
and treat this component as a ``nuisance parameter'' to be
marginalised over, and concentrate on the cosmological information in
the linear part of Eq.~(\ref{eq:pk_s1}).

For linearly evolving density fields, if we can write down a linear
mass conservation equation $\theta=-a\dot{\delta_m}$, then the growth
factor of $\theta$, $D_\theta\propto a\dot{D}$. Defining $f\equiv
d\ln D/d\ln a=\dot{D}/(DH)$, if $\delta_g$ is perfectly correlated
with $\theta$ everywhere, then the power spectra in
Eq.~(\ref{eq:pk_s1}) have the same shape, and we have that
\begin{equation}
 P_g^s({\bf k}) = P_g({\bf k}) \left[1
                        + 2\mu^2\beta
                        + \mu^4\beta^2\right]
                      F\left(\frac{k^2\mu^2\sigma_v^2}{H^2(z)}\right),
 \label{eq:pk_s2}
\end{equation}
where $\beta\equiv f/b$. 

Note that writing Eq.~(\ref{eq:pk_s2}) in terms of $\beta$ suggests
that the large-scale redshift-space distortions depend on the galaxy
bias. We expect the motion of galaxies to locally match those of the
matter field, so two galaxies at the same location would have the same
peculiar velocities irrespective of their internal properties: they
simply act as test particles within the matter flow. This assumption
is part of Eq.~(\ref{eq:pk_s1}): the dependence on bias simply comes
from expecting a multiplicative correction to the observed galaxy
power spectrum for the redshift-space distortions. If we instead
consider modelling an additive contribution, then the dependence on
bias can be broken.

\begin{figure}[t]
 \begin{center}
 \epsfysize=3.0truein
 \epsfxsize=3.0truein
   \epsffile{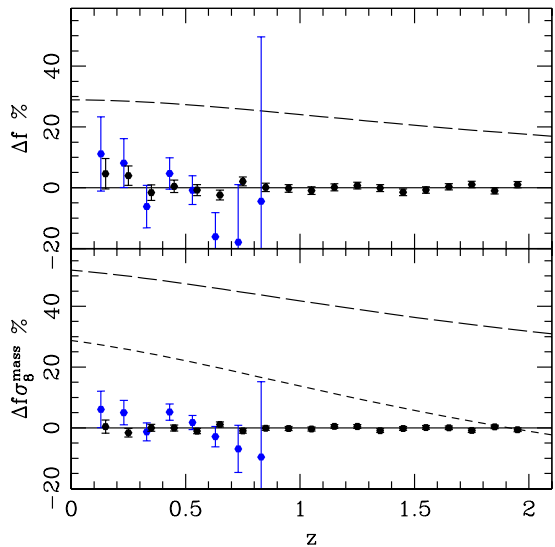}
   \caption{\footnotesize The top panel shows the percentage
     difference in $f$ between the sDE and DGP model (long dashed
     line). The sDE and DGP models are described in detail in
     Section~\ref{sec:mods}. The background expansion has been matched
     between these two models. The bottom panel shows the percentage
     difference of $f\sigma_8^{\rm mass}$ between sDE and DGP. Here,
     the long-dashed curve includes CMB data, normalising the models
     at the epoch of last scattering (using $\Delta^2_{\zeta_{ini}}$),
     while the dashed curve shows the model normalised using a low
     redshift measurement of $\sigma_8^{\rm mass}(z=0)=0.82$, matching
     the 5-year WMAP best-fit $\Lambda$CDM value
     \cite{Komatsu08}. The blue and black error bars are estimated
     from BOSS and EUCLID respectively (see
     Section~\ref{sec:future_constraints} for details).}
\label{fig:f_v}
\end{center}
\end{figure}

\begin{figure}[t]
 \begin{center}
 \epsfysize=3.0truein
 \epsfxsize=3.0truein
   \epsffile{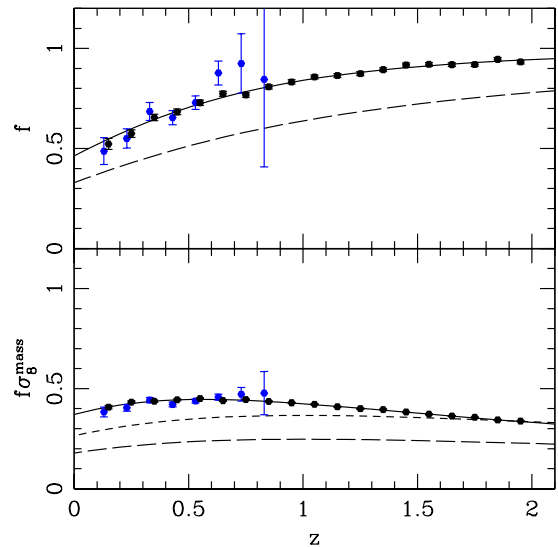}
   \caption{\footnotesize The top panel shows the time evolution of
     $f$: solid curve is for sDE model, and a long-dash curve is for
     DGP. The sDE and DGP models are described in detail in
     Section~\ref{sec:mods}. The bottom panel shows the evolution of
     $f\sigma_8^{\rm mass}$: the solid curve shows the sDE model, the
     long-dash curve DGP normalized using $\Delta^2_{\zeta_{ini}}$,
     and the short-dashed line, DGP $\sigma_8^{\rm
       mass}(z=0)=0.82$. The blue and black error bars are estimated
     from BOSS and EUCLID respectively (see
     Section~\ref{sec:future_constraints} for details).}
\label{fig:f_fsig8}
\end{center}
\end{figure}

Comparing Eq.~(\ref{eq:pk_s1}) and (\ref{eq:pk_s2}) shows that,
although many observational studies present results in terms of
$\beta$, the fundamental constraint from the $\mu$ dependence of the
normalisation of Eq.~(\ref{eq:pk_s1}) is on the normalisation of
$P_{\theta\theta}$, and the cross power $P_{g\theta}$. If we assume
that the continuity equation holds, and we have a local linear bias so
$\delta_g=b\delta_{\rm mass}$, then the normalisation of $P_{gg}$
depends on $(b\sigma_8^{\rm mass})^2$, $P_{g\theta}$ depends on
$(f\sigma_8^{\rm mass})(b\sigma_8^{\rm mass})$, and $P_{\theta\theta}$
depends on $(f\sigma_8^{\rm mass})^2$. If the redshift-space
distortions are modeled as a multiplicative component, then we
constrain ($b\sigma_8$,$\beta$), while if they are modeled as an
additive component we measure ($b\sigma_8$,$f\sigma_8$).  These
constraints will be a simple transform of each other, and should give
the same likelihood surface if we transform into the same basis (this
will be considered further in future work\cite{song09}). Previous
analyses have, in general, converted observational results into a
constraint on $f$ using a separate measurement of bias
(e.g. \cite{guzzo08}). A perfect estimate of bias would allow us to
break $f\sigma_8^{\rm mass}$ and $b\sigma_8^{\rm mass}$ to provide
separate constraints on $\sigma_8^{\rm mass}$, $f$ and $b$. Note that
$f$, $\sigma_8^{\rm mass}$ and $b$ are all redshift dependent
functions.

In Fig.~\ref{fig:f_v} and Fig.~\ref{fig:f_fsig8}, we show that the
combined quantity, $f\sigma_8^{\rm mass}$ is sufficient to distinguish
between DGP and sDE models described in Section~\ref{sec:mods}, if the
fluctuations are additionally normalised using CMB data. If however,
we normalise the model predictions for $f\sigma_8^{\rm mass}$ using a
present day measurement of $\sigma_8^{\rm mass}$, then the predicted
differences in $f$ and $f\sigma_8^{\rm mass}$ between the DGP and sDE
models are very similar, a consequence of the weak difference in the
low-redshift evolution of $\sigma_8^{\rm mass}$. Normalising the model
prediction for $f\sigma_8^{\rm mass}$ at low redshift is a fairer
comparison between using the two ways of describing the redshift-space
distortion constraints, as this effectively only uses the ratios
between measurements of $f\sigma_8^{\rm mass}$ at different
redshifts. 

Given that no technique is significantly better at distinguishing these
models in terms of the statistical error, it is worth considering the
potential for including systematic errors using both techniques. Here,
$f\sigma_8^{\rm mass}$ wins, since it can be measured without knowing
the bias $b$ or the amplitude of the matter fluctuations
$\sigma_8^{\rm mass}$.  

In this analysis we have compared two models of acceleration. Clearly,
future measurements will be used to test a wide range of models. It is
clear that for some of these $f$ will be superior, while for others
using $f\sigma_8^{\rm mass}$ directly would provide tighter
constraints. In particular, using $f\sigma_8^{\rm mass}$ directly can
be advantageous for a model such as the DGP model tested here in which
growth of structure formation is suppressed compared with
$\Lambda$CDM. This is not true for other theoretical models in which
growth of structure is less suppressed, e.g. normal branch DGP,
compared with the self-accelerating DGP model considered here
\cite{song07}.

\subsection{current constraints}  \label{sec:current}

Redshift-space distortions have recently been measured at $z=0.77$
using the VIMOS-VLT Deep Survey (VVDS)\cite{guzzo08}. They use
$\sigma_8^{\rm mass}(z=0) = 0.78\pm0.03$ measured by the WMAP
experiment, extrapolated to the survey redshift $z=0.77$
$\sigma_8^{\rm mass}(z=0.77) = 0.54\pm0.03$ (assuming a fiducial
$\Omega_m=0.25$, $\Omega_\Lambda=0.75$ cosmology) to measure the
bias. Using this estimate, they convert their measured value of
$\beta$ to provide a constraint on $f(z)$. If we do not do this bias
modelling, then we can reconstruct the constraint on $f\sigma_8^{\rm
  mass}$ from their quoted results. To do this, we take their
$f(z=0.77)=0.91\pm0.36$ measurement and multiply by the extrapolated
WMAP result $\sigma_8^{\rm mass}(z=0.77)=0.54$ Doing this, their
redshift-space distortion measurement becomes $f\sigma_8^{\rm
  mass}=0.49\pm0.18$ at $z=0.77$. We have conservatively assumed that
all of the quoted error comes from the LSS observations rather than
the extrapolated CMB normalisation and subsequent bias estimate.

At lower redshift, we use results from a spherical harmonics analysis
of the 2dFGRS\cite{percival04}. This work did not provide a direct
constraint on $f\sigma_8^{\rm mass}$ at the fiducial redshift of the
2dFGRS, but extrapolated this to $z=0$. Without loss of generality, we
can undo this extrapolation using their fiducial cosmology. For this
cosmology ($\Omega_m=0.3$, $\Omega_\Lambda=0.7$),
$D(z=0.17)/D(z=0)=0.916$, and $f(z=0.17)/f(z=0)=1.2$. Their
extrapolated $z=0$ measurement of $f\sigma_8^{\rm mass}=0.46\pm0.06$
should therefore be considered a measurement $f\sigma_8^{\rm
  mass}=0.51\pm0.06$ at $z=0.17$.

We also include a measurement of redshift-space distortions from the
SDSS LRG catalogue\cite{tegmark06}. This work measured
$\beta=0.309\pm0.035$ at $z=0.35$. To convert to a constraint on
$f\sigma_8^{\rm mass}$, we multiply by $b\sigma_8^{\rm mass}=1.43$
from the joint SDSS+CMB fit presented. Here we assume that, because of
the wide survey geometry in the SDSS, the measurement error on $\beta$
is significantly larger than that on $b\sigma_8^{\rm mass}$. This
gives a measurement $f\sigma_8^{\rm mass}=0.44\pm0.05$.

\subsection{future constraints} \label{sec:future_constraints}

In order to see how these constraints will improve with the next
generation of experiments, we have calculated expected errors for
three types of future experiment. We have considered a ground based
redshift-survey of galaxies out to $z=0.8$, based on the BOSS
experiment, part of the SDSS-III (see {\tt www.sdss3.org}). For this
survey we have assumed that redshifts will be measured for
$1.5\times10^6$ galaxies over $10\,000\,{\rm deg}^2$ with
approximately constant number density out to $z=0.7$. We also consider
a next generation space based mission that could measure
$0.5\times10^9$ galaxies in a volume $V=1\times 10^{11}\,h^{-3}\,{\rm
  Mpc}^3$ out to $z=2$. Such a survey is proposed has been proposed to
ESA mission as part of the EUCLID mission, and is the result of the
merging between SPACE and DUNE \cite{roberto07,cimatti08,dune}. 
We have also considered the type of
experiment that could be accomplished using a wide-field multi-object
spectrograph on a 8m-class telescope. Here, we took the ``fiducial
survey'' parameters presented for the proposed WFMOS instrument
\cite{glazebrook05}. These are two surveys with number density
$5\times10^{-4}\,h^3\,{\rm Mpc}^{-3}$, one of $2\,000\,000$ galaxies
with $0.5<z<1.3$, and one of $600\,000$ galaxies with $2.3<z<3.3$. The
numbers and the derived expected measurements for all of these surveys
should provide an approximate guide to the improvements that are
expected beyond current observations.

To calculate expected errors for these surveys, we use the fitting
formula presented in \cite{guzzo08}, which was derived from numerical
simulations. This formula was presented as providing the error on $f$,
and relies on having an independent accurate measurement of
$\sigma_8^{\rm mass}$ (which is itself dependent on cosmology through
the growth rate, rather than its derivative). In the above discussion
we argued that it is more consistent to measure and test a constraint
on $f\sigma_8^{\rm mass}$, and we translate the \cite{guzzo08} formula
as giving a fractional error
\begin{equation}
  \frac{\Delta(f\sigma_8^{\rm mass})}{f\sigma_8^{\rm mass}}
  = \frac{50}{\left(0.2*\langle n_g \rangle\right)^{0.44} \sqrt{V}},
\end{equation}
where $V$ is the volume of the survey (or part of the survey under
consideration) in $(\mpcoh)^3$, and $n_g$ is the galaxy number density
in $(\hompc)^3$. By performing this translation we have conservatively
assumed that the error predicted on $f$ by \cite{guzzo08} was
dominated by the measurement of $f\sigma_8^{\rm mass}$, rather than
the translation to $f$. The \cite{guzzo08} formula has behaviour very
close to Poisson with $\propto1/N_{\rm gal}$, consistent with the idea
that we are not concerned with the scales on which we see galaxy
pairs, but simply want amplitude information so all pairs count
equally, and also assumes that we can recover information from both
$P_{g\theta}$ and $P_{\theta\theta}$~\cite{white08}.  This formulae is
too simplistic to capture many of the dependencies, such as on galaxy
bias and power spectrum shape, so we have performed a full Fisher
matrix calculation.  This validates this simple \cite{guzzo08}
formulae for reasonable values of galaxy bias, and galaxy power
spectra. As our aim in this paper is to present the case for
redshift-space distortions as a probe of dark energy models, the
\cite{guzzo08} formula is adequate for our purpose. We convert to
calculate errors on $f$ for Figs.~\ref{fig:f_v} and~\ref{fig:f_fsig8}
assuming that
\begin{equation}
  \frac{\Delta f}{f}=\sqrt{\left(\frac{\Delta
  f\sigma_8}{f\sigma_8}\right)^2+\left(\frac{\Delta
  \sigma_8}{\sigma_8}\right)^2}\,, 
\end{equation} 
with $\sigma_8=0.82\pm0.03$, following the $\Lambda$CDM constraint
from the 5-year WMAP data \cite{Komatsu08}, and assume that the same
level of accuracy can be achieved at low redshift by the next
generation of lensing experiments.

\subsection{model comparison}  \label{sec:mods}

\begin{figure*}[t]
\centerline{
\epsfxsize=6.3truein\epsffile{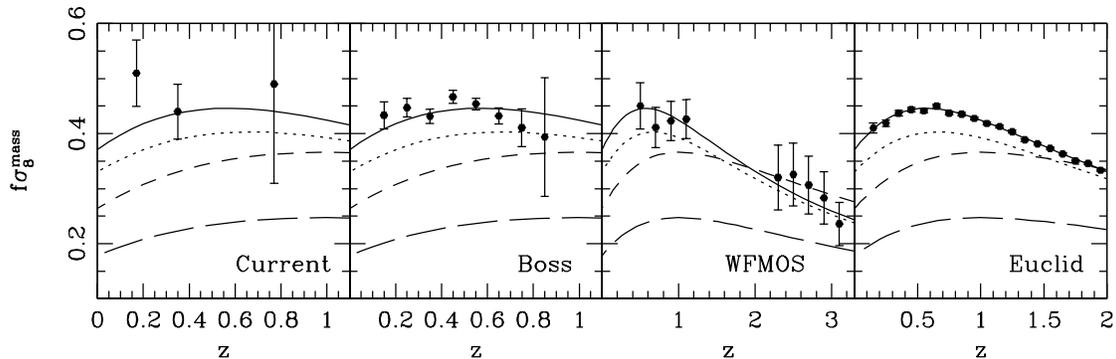}$\,$}
\caption{ The solid curve represents sDE, long-dash curve represents
  DGP, and dotted curve represents IDE. These models predict
  approximately the same background expansion, which is normalised at
  high redshift to match the fluctuations observed in the CMB. The
  left panel shows the current constraints discussed in the text. The
  three other panels show simulated data for BOSS, WFMOS and
  EUCLID-type experiments (see text for details).}
\label{fig:sigma8_theta}
\end{figure*}

Current and future constraints on $f\sigma_8^{\rm mass}$ are compared
with various models in Fig.~\ref{fig:sigma8_theta}. The models are all
normalised to match the amplitude of fluctuations seen in the CMB by
the 5-year WMAP experiment \cite{Komatsu08}. We see that future
experiments will provide an order-of-magnitude improvement over
current studies, and will provide constraints over a range of redshift
from a single survey.

When comparing against models, we have not marginalised over
parameters that are determined by the background, but instead fix
parameters close to our sDE fiducial model in fitting to current
geometrical constraint given by combination of WMAP, SN and Hubble key
project. Fig.~\ref{fig:sigma8_theta} therefore shows the differences
between structure growth in the different models, rather than
performing a full Likelihood test given current data. It shows what we
can tell about structure formation if we have perfect information
about the background expansion.

In detail, for DGP, we fix the distance to recombination at $z_{\rm
  lss}=1088^{+1}_{-2}$ through the measurement of the acoustic peak
scale $l_{A}=302^{+0.9}_{1.4}$ and its length calibration through the
matter density $\Omega_m h^2=0.128\pm 0.008$, and we fit the DGP
cosmological parameter to the combination of SN and $H_0$ from Hubble
Key Project. Open curvature is introduced to enhance the fit to data
set. For cDE, we use the effective dark energy which makes the cDE
geometrical factor exactly identical to the open DGP model used
here. For IDE, we fix $\Omega_m h^2$ at its best value measured by
WMAP, and we allow $H_0$ to be varied within 1-$\sigma$ range of
measured $H_0=72\pm 8 {\rm km/sec/Mpc}$. Here we choose $H_0$ at the
edge of upper bound, $H_0=80 {\rm km/sec/Mpc}$, at $\gamma=0.3$. If
$\gamma$ becomes bigger than this value, then it is expected that it
deviates significantly from the geometrical measurement as well as
structural departure considered in this section.

In this paper we have followed standard convention and used
$\sigma_8^{\rm mass}$ to normalise the matter power spectrum. The
redshift-space distortion measurements actually constrain fluctuations
on a range of scales larger than those probed by $\sigma_8$, but this
does not matter if we have a constant power spectrum shape. If the
power spectrum shape is not constant, then the statistic used would
have to be revised, and the full modelling of $P(k)$ included. This is
beyond the scope of the current paper, and we assume a constant $P(k)$
shape. Here, we have assumed that the power spectrum of the initial
fluctuations in the comoving gauge as measured by WMAP, has amplitude
$\Delta_{\zeta_{\rm ini}}^2=2.4\times 10^{-9}$. We assume that the
time-dependent growth function of $\Phi$ is normalized to unity at the
onset of matter domination and that the scale dependent transfer
function is normalized to unity in the $k\rightarrow 0$ limit. For DGP
and cDE models, we tune the acoustic peak structure to be identical
to sDE with same $\omega_m$ and $\omega_b$ and with fixing angular
diameter distance to last scattering surface. For IDE, we do the same
with the extra assumption that the earlier DE decay is
negligible. This is a good approximation for small coupling limit used
in this paper. The full treatment of the CMB in IDE models is beyond
the scope of the current paper.

\section{further applications of peculiar velocity measurements}

\label{sec:further_tests}

We now discuss additional cosmological applications that can be
developed from peculiar velocity measurements. This is a development
of previous work that considered constraints from observations of
projected power spectra~\cite{song08a}. Here we assume that we have
measurements of the growth of $P_{\theta\theta}$. We have a similar
measurement from the combination of $P_{g\theta}$ and $P_{gg}$, but as
this uses the continuity equation for baryons we treat this
separately. First, we consider how a redshift dependent measurement of
the amplitude of $P_{\theta\theta}$ can be used to reconstruct
$\Psi$. We quantify this normalisation using $\sigma_8^{\theta}$,
defined as the rms fluctuations of $\theta$ averaged over spheres of
radius $8\mpcoh$.

If we are able to probe the history of $\sigma_8^{\rm mass}$, in
addition to $\sigma_8^\theta$, then we can complete the structure
formation test. Measuring the derivative of $\sigma_8^{\rm mass}$ with
respect to time would enable us to test the continuity equation by
comparing it with $\sigma_8^\theta$.  This test eliminates many
different theoretical models explaining cosmic acceleration, and is
potentially more powerful than simply measuring only one of these
quantities. Finally, the reconstructed $\Psi$ created from
$\sigma_8^\theta$ can be compared with $\Phi$ estimated from
$\sigma_8^{\rm mass}$, which will constrain the anisotropic stress.

Measuring the evolution of $\sigma_8^{\rm mass}$ is problematic
because we cannot observe the mass directly. We could try to predict
how galaxies trace the matter
(e.g. \cite{BBKS,cole89,seljak00,peacock00,cooray02}), or could
measure this bias, for example using higher order statistics
(e.g. \cite{verde02}). Alternatively, we could use weak lensing
measurements \cite{amendola08a,jain07,acquaviva}. We expect that these
methods will give constraints on bias at the percent level, by the
time that redshift-space distortion measurements are obtained from the
EUCLID experiment. Note that the weak-lensing constraints could
themselves come from an imaging component of the EUCLID mission. For
our proposed tests that require bias measurements, we therefore assume
an expected uncertainty of $\Delta b/b \sim 0.02$. The models used
match those described in Section~\ref{sec:mods}.

\subsection{Reconstruction of perturbed potential}
\label{sec:get_potential}

If $\sigma_8^\theta$ can be measured with sufficient accuracy that its
derivative can also be determined, then we can estimate $\Psi$ from
Eq.~(\ref{eq:Euler_X}). The measured derivative of $\sigma_8^\theta(z)$
as a function of redshift can be written
\begin{equation}
\frac{d\sigma_{8}^{\theta\,i}}{dt}
  =-\frac{H^i}{a}
\frac{\sigma_{8}^{\theta\,i+1}-\sigma_{8}^{\theta\,i-1}}{2\Delta z}\,,
\end{equation}
where $i$ denotes each redshift bin. From this, we can construct an
estimator of $\hat\Psi^i$ from
\begin{equation}
k^2\hat\Psi^i=
-H^i\frac{\sigma_{8}^{\theta\,i+1}-\sigma_{8}^{\theta\,i-1}}{2\Delta z}
+\frac{H^i}{1+z^i}\sigma_{8}^{\theta\,i}\,.
\end{equation}
We show estimates of how well future experiments will be able to
reconstruct $\Psi$ using this method in the top panel of
Fig.~\ref{fig:LSS_test}.  If there is no interaction between dark
matter and dark energy, then $\langle\hat\Psi\rangle=\Psi$, since
there is no change from the standard Euler equation. This offers a new
way to probe the geometrical perturbation other than weak lensing
experiment. Even with weak lensing measurements, it is interesting to
have $\Psi$ separately, because weak lensing probes the combination
$\Phi-\Psi$.

\subsection{Test on continuity equation}\label{subsec:cont}

\begin{figure}[htbp]
 \begin{center}
 \epsfysize=4.5truein
 \epsfxsize=3.truein
   \epsffile{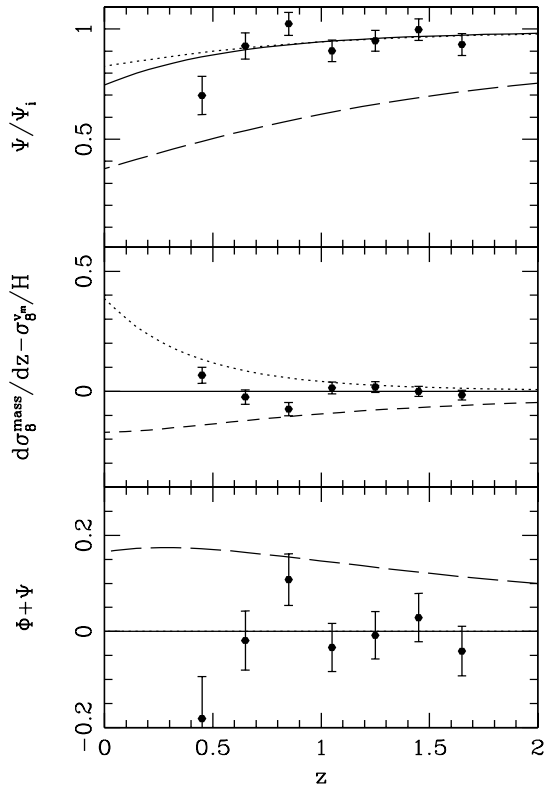}
   \caption{\footnotesize Constraints on physical measurements that
     can be derived from a EUCLID type survey. The solid lines,
     long-dash lines, dashed lines and dotted lines are sDE models,
     DGP, cDE model and IDE model respectively. The top panel shows
     the reconstruction of $\Psi$ from $\sigma_{8}^{\theta}$, the
     middle panel shows the test of the continuity
     equation~\cite{song08b}, and the bottom panel shows the
     constraint on anisotropic stress. The errors shown in this figure
     assume that we can recover $\sigma_{8}^{\theta}$ from both
     $P_{\theta\theta}$ and $P_{g\theta}$.}
\label{fig:LSS_test}
\end{center}
\end{figure}

If we can measure the evolution of $\sigma_{8}^{\rm mass}$ and
$\sigma_{8}^{\theta}$ simultaneously, then the continuity equation can
be tested.  The measured $\sigma_8^{\rm mass}$ at each redshift bin
leads us to estimate its derivative in terms of time, which is defined
by,
\begin{equation}
  \frac{\sigma_{8}^{\delta\,i+1}-\sigma_{8}^{\delta\,i-1}}{2\Delta z}
  =\frac{\sigma_{8}^{\theta\,i}}{H_i}\,,
\end{equation}
where $i$ denotes the number of redshift bin.  In the middle panel of
Fig.~\ref{fig:LSS_test}, we show the estimated errors in constraining
the continuity equation with EUCLID.  The departure from the
continuity equation of cDE model having the identical structure
formation to DGP model (dash curve) is detectable.  It means that MG
models can be distinguishable from DE type model, i.e. modified
gravity is detectable.  The dotted curve representing IDE has
detectable departure from the continuity equation.  This test will be
a crucial future test for exotic DE models, but requires either galaxy
bias to be fully understood or modelled. An alternative test using
weak-lensing observations is possible.

\subsection{Constraint on anisotropic stress}

With the same assumption as in the previous subsection of an
independent measurement on bias, we can extend the analysis to
constrain anisotropic stress.  The curvature perturbations are
determined by
\begin{equation}
 k^2\Phi^i=4\pi G_N a^2 \rho_m\delta_m^i\,,
\end{equation}
so a measurement of the normalisation of the density field leads to a
constraint on $\Phi^i$. Using this measurement and the measurement in
Section~\ref{sec:get_potential}, we are able to constrain anisotropic
stress by comparing reconstructed $\Psi^i$ and $\Phi^i$. 
We show in the bottom panel of Fig.~\ref{fig:LSS_test}, that
we are able to constrain the level of anisotropic stress predicted
from DGP model.

\section{conclusions}

One of the best ways of measuring this structure growth is to use
observations of redshift-space distortions. This is an interesting and
timely subject given current observations and those planned on a
10--20 year timescale \cite{Wang08,sapone07,guzzo08}. In this paper we
have reviewed the importance of redshift-space distortion measurements
given that they provide a measurement of structure growth that is
independent of galaxy density bias. We have argued that peculiar
velocity measurements are best presented in terms of
$\sigma_{8}^{\theta}$, or $f\sigma_8^{\rm mass}$ for models where the
continuity equation holds. The independence from galaxy density bias
has not been widely covered in previous literature. Most previous
analyses have considered measuring $\beta$ and $b\sigma_8^{\rm mass}$,
although there are some exceptions\cite{percival04}. Although
extremely simple, we have focused on the density bias independent
constraint resulting from multiplying these together. The physical
origin of such a constraint is that linear velocities, which scale
with the derivative of the growth factor, depend only on the matter
velocity field.

The primary conclusion from our work is that constraints on
$\sigma_{8}^{\theta}$ or $f\sigma_8^{\rm mass}$, are extremely good at
helping to distinguish between the dark energy models that we reviewed
in Section~\ref{sec:models}. In fact, as we show in
Fig.~\ref{fig:f_v}, these constraints are equivalent to similar
percent measurements of $f$ for some models of cosmic
acceleration. They also have the simplicity of not having to model
galaxy bias. The simple formula of \cite{guzzo08} has been adapted to
determine constraints on $f\sigma_8^{\rm mass}$, and shows future
constraints in Fig.~\ref{fig:sigma8_theta}. Although the
\cite{guzzo08} formula is simplistic, and might not be believed at the
percent level, it shows that we can expect a huge step forwards in
redshift-space distortion measurements with the next generation of
surveys.

Going beyond simply obtaining a single measurement of
$\sigma_{8}^{\theta}$ or $f\sigma_8^{\rm mass}$, we have considered
how the underlying perturbation evolution can be tested using peculiar
velocity measurements. Peculiar velocity measurements are important
because they can be used to reconstruct Newtonian potential $\Psi$
which sources the dynamics of a galaxy given by Euler
equation. Weak-lensing only measures $\Psi$ in the combination
$\Phi-\Psi$, so redshift-space distortions offer a complementary test
of perturbations.

We have considered how peculiar velocities can be used to test the
continuity equation, which is worthwhile since there are many
theoretical models which fail to satisfy this relation. If dark energy
couples to matter, then current constraints show that it must couple
to the dark matter and not to baryonic material. The coupling of dark
energy to dark matter modifies the Euler equation for dark matter, and
breaks the equivalence principle between dark matter and baryon. This
difference in free-fall breaks the continuity equation in which the
peculiar velocity of matter is estimated using baryons, while we
consider the growth of fluctuations in all matter. In addition, this
test can tell if there is dark energy clustering which deepens the
curvature potential well because we measure the peculiar velocity of
the matter not of the total energy density. Finally we have considered
how we can constrain the anisotropic stress by comparing $\Phi$ and
$\Psi$, reconstructed from the density fields and peculiar velocity
respectively.

\section*{Acknowledgments}

We would like to thank Luca Amendola, David Bacon, Luigi Guzzo 
and Martin White for useful
conversations. Y-SS and WJP are grateful to support from STFC.


\end{document}